\documentclass[12pt,a4paper,final]{iopart}

\usepackage{iopams}
\usepackage{graphicx}
\usepackage[breaklinks=true,colorlinks=true,linkcolor=blue,urlcolor=blue,citecolor=blue]{hyperref}
\usepackage[font=footnotesize,labelfont=bf,
   justification=justified,
   format=plain]{caption}

\begin{document}

\title[]{Weak Electronic Correlations Observed in Magnetic Weyl Semimetal Mn$_3$Ge}

\author{Susmita Changdar$^{1,*}$, Susanta Ghosh$^{1,*}$, Anumita Bose$^2$, Indrani Kar$^1$, Achintya Low$^1$, Patrick Le Fevre$^3$, Fran\c{c}ois Bertran$^3$, Awadhesh Narayan$^2$, Setti Thirupathaiah$^{1,\dag}$}
\address{$^1$Condensed Matter and Materials Physics Department, S. N. Bose National Centre for Basic Sciences, Kolkata, West Bengal-700106, India}
\address{$^2$Solid State and Structural Chemistry Unit, Indian Institute of Science, Bangalore 560012, India}
\address{$^3$SOLEIL Synchrotron, L'Orme des Merisiers, D\'epartementale 128, F-91190 Saint-Aubin, France}
\address{$^*$ These authors contributed equally}
\address{$^{\dag}$ Corresponding author: setti@bose.res.in}

\begin{indented}
\item[]\today
\end{indented}

\begin{abstract}
Using angle-resolved photoemission spectroscopy (ARPES) and density functional theory (DFT) calculations, we systematically studied the electronic band structure of Mn$_3$Ge in the vicinity of the Fermi level. We observe several bands crossing the Fermi level, confirming the metallic nature of the studied system. We further observe several flat bands along various high symmetry directions, consistent with the DFT calculations. The calculated partial density of states (PDOS) suggests a dominant Mn $3d$ orbital contribution to the total valence band DOS. With the help of orbital-resolved band structure calculations, we qualitatively identify the orbital information of the experimentally obtained band dispersions. Out-of-plane electronic band dispersions are explored by measuring the ARPES data at various photon energies. Importantly, our study suggests relatively weaker electronic correlations in Mn$_3$Ge compared to Mn$_3$Sn.
\end{abstract}

 \vspace{2pc}

\section{Introduction}
The coexistence of magnetism and topology in magnetic semimetals manifests several intriguing quantum materials in the condensed matter such as the Weyl semimetals~\cite{Burkov2011, Wan2011, doi:10.1146, RevModPhys.90.015001}. Mn$_3$Ge is believed to be one such magnetic Weyl semimetal from the family of Mn$_3$X (X= Sn \& Ge) kagome systems,  with noncollinear antiferromagnetic (AFM) order, demonstrating exotic electronic properties like giant anomalous Hall effect (AHE)~\cite{Nayak2016,PhysRevApplied.5.064009,nakatsuji2015large,PhysRevB.101.094404}, anomalous Nernst effect~\cite{narita2020effect,narita2017anomalous,PhysRevB.100.085111,PhysRevB.96.224415,PhysRevLett.119.056601}, Kerr effect~\cite{wu2020magneto,higo2018large,doi:10.1063}, and flat bands along with Weyl nodes near the Fermi level~\cite{yang2017topological,chen2021anomalous},  leading to compelling quantum technological application~\cite{Popovic2004, Ramsden2006}. Usually, the Weyl points are manifested either by the broken time-reversal symmetry (TRS) or by the broken inversion symmetry. Out of these two categories, the TRS broken systems are in principle magnetic WSMs which host strong electronic correlations~\cite{pan2022giant,destraz2020,ikhlas2022}, exhibiting several fascinating bulk electronic and magnetic properties. On the other hand, in the case of the Mn$_3$X family of materials, it is the translational plus time-reversal symmetry (TRS) breaking that gives rise to the Weyl physics~\cite{liu2017,chen2021anomalous,PhysRevB.96.224415,PhysRevB.106.195114}. An inversion symmetry broken (under the time-reversal symmetry) Weyl system can host a minimum of two pairs of Weyl nodes (four Weyl points), while a time-reversal symmetry broken (under the inversion symmetry) Weyl system can host a minimum of one pair of Weyl nodes (two Weyl points).  Each pair of Weyl nodes with opposite chirality acts as the source and sink of the Berry curvature, leading to large AHE~\cite{Kübler_2014,Nayak2016,PhysRevApplied.5.064009,wang2021}. Thus, a systematic study of the low-energy electronic band structure of these systems is crucial for understanding the observed large AHE.

\begin{figure*}[ht!]
	\includegraphics [width=\linewidth]{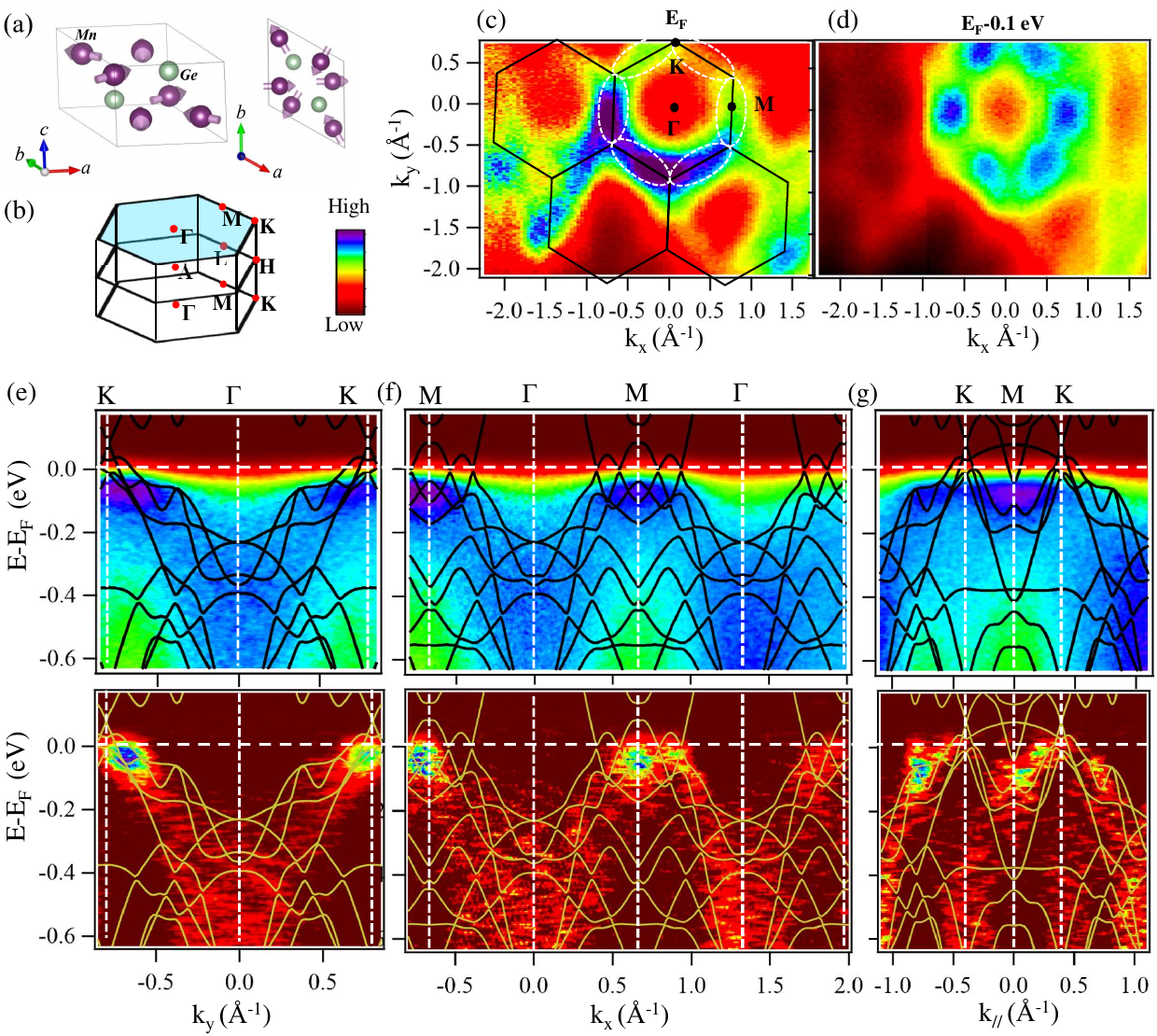}
	\caption{Hexagonal (a) crystal structure and (b) Brillouin zone of Mn$_3$Ge. (c) In-plane Fermi surface map. (d) Constant energy contour map taken at 0.1 eV below the Fermi level. Energy distribution maps (EDMs) taken along (e) $\Gamma-K$, (f) $\Gamma-M$, and (g) $M-K$ high symmetry directions are shown on the top-panel. The bottom panels of (e)-(g) demonstrate the second derivatives of the corresponding EDMs. The overlapped black-solid curves in (e)–(g) are the DFT band structure.}
	\label{fig1}
\end{figure*}


Although there exist many density functional theory (DFT) calculations exploring the topological properties of Mn$_3$X (X = Sn and Ge ) systems by studying their electronic band structure near the Fermi level~\cite{yang2017topological, Kübler_2017, PhysRevB.102.094403, xu2020high,elcoro2021magnetic}, so far only one ARPES report is available experimentally demonstrating the electronic band structure of Mn$_3$Sn~\cite{kuroda2017}. The DFT calculations predict several Weyl points near the Fermi level from both Mn$_3$Sn and Mn$_3$Ge. A recent theoretical report even suggested that the Weyl points in these systems are accidental~\cite{Bernevig2022}. However, experimentally, it is hard to detect these Weyl points due to difficulties in resolving the valence band electronic structure. Because of this complexity, not much excitement is generated in the research community for experimentally studying the electronic band structure of these very fascinating magnetic topological Weyl semimetals.

In this manuscript, we present the low-energy electronic band structure of Mn$_3$Ge studied by using the ARPES technique and compared it with the calculated band structure using DFT.  We find a qualitative agreement between the ARPES data and DFT calculations.  However, detection of the theoretically predicted Weyl points from our ARPES data has been quite challenging as they are predicted well above the Fermi level~\cite{yang2017topological,chen2021anomalous}. Since Mn$_3$X compounds are suggested to show strong electronic correlations~\cite{kuroda2017,PhysRevB.106.205103}, instead of searching for the Weyl points,  we focused on the details of low-energy electronic band structure along the in-plane and the out-of-plane ($k_z$) orientations. The partial density of state (PDOS) calculations confirm the presence of dominant Mn $3d$ orbital characters near the Fermi level. We further carried out the orbital-resolved DFT calculations and compared them with the ARPES data to understand the orbital contributions to the experimental electronic band structure. In addition, as the Mn$_3$Ge is constructed from layered kagome lattice, these are predicted to host flat bands~\cite{li2021dirac,kang2020,kang2020dirac}. Our ARPES data demonstrate the presence of such flat bands along different high-symmetry directions of the Brillouin zone.

\section{Experimental Details}
Mn$_3$Ge single crystals were grown by the melt-growth method~\cite{Nayak2016}. To grow the crystals, stoichiometric amounts of high quality Mn (99.95\%) and Ge (99.999\%) powders were mixed thoroughly under an argon environment and sealed in a quartz tube at 10$^{-4}$ mbar vacuum. The quartz tube was kept at 1050$^o$C for 24 hours before cooling down to 740$^o$C at a rate of 2$^o$/hr. After prolonged annealing for 120 hrs, the tube was quenched in ice water. In this way, we could grow the single crystals of Mn$_3$Ge with a typical size of 2$\times$3 mm$^2$. As grown single crystals were studied structurally using the X-ray diffraction (XRD) technique to confirm the hexagonal crystal structure with a space group of P6$_{3}$/mmc (194). The energy dispersive X-ray spectroscopy (EDS) study suggests an actual chemical composition of Mn$_{2.94\pm0.03}$Ge for the single crystals which were used for the ARPES measurements. The ARPES measurements were performed at CASSIOP\'EE beamline in the SOLEIL synchrotron radiation centre, France, equipped with a SCIENTA R4000 analyzer. The samples were cleaved $in-situ$ under a chamber vacuum of $\approx 5\times10^{-11}$ mbar. The data were collected using photon energies between 83 and 143 eV at a sample temperature of 35 K. The total energy resolution was set to 15 meV.

\section{Computational Methods}
Band structure calculations were performed on Mn$_3$Ge using density functional theory (DFT) within the generalized gradient approximation (GGA) of Perdew, Burke, and Ernzerhof (PBE) exchange and correlation potential~\cite{Perdew1996} as implemented in the Quantum Espresso simulation package~\cite{Giannozzi2009, Giannozzi2017}. Ultrasoft non-relativistic and fully relativistic pseudopotentials were used to perform the calculations without spin-orbit coupling (SOC) and with SOC, respectively. The electronic wave function is expanded using plane waves up to a cutoff energy of 50 Ry. Brillouin zone sampling is done over a $\Gamma$-centred 7$\times$7$\times$7 Monkhorst-Pack k-grid. The internal coordinates of the system are relaxed before producing the band structure.  The ground state of this system is in the noncollinear AFM order with each Mn magnetic moment aligned at an angle of 120$^o$ with its neighbouring moment vector as shown in Fig.~\ref{fig1}(a). Each Mn atom in Mn$_3$Ge carries an average magnetic moment of 2.638 $\mu_B$. Also, a slight tilt in Mn moments causes an in-plane net magnetic moment of $\approx$ 0.008 $\mu_B$/cell. We observe that overall the experimental band structure matches better with DFT calculations for $U=0~eV$, consistent with previous studies~\cite{yang2017topological, kuroda2017}.

\section{Results and Discussion}
\begin{figure*}
	\includegraphics[width=\linewidth]{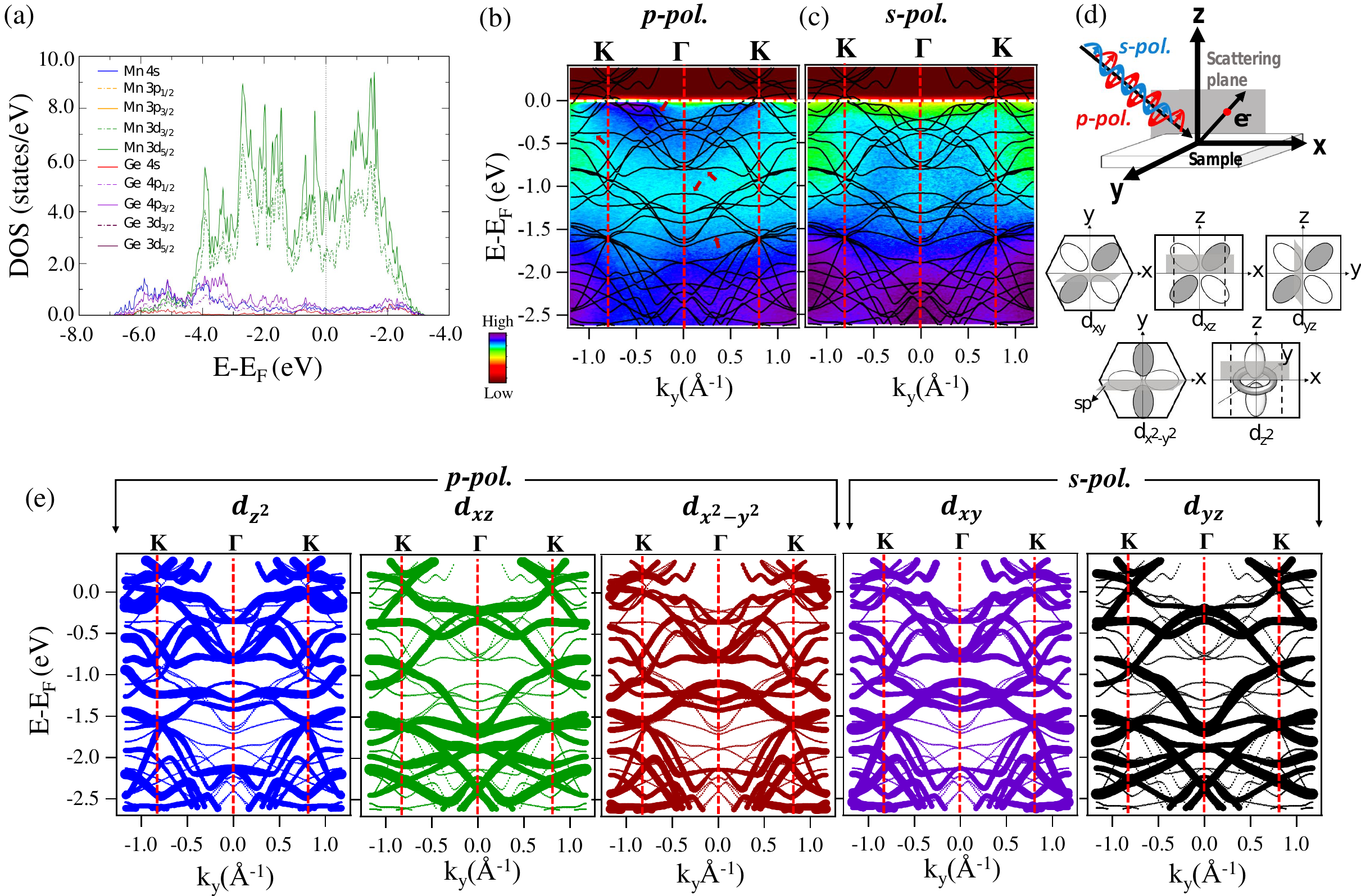}
	\caption{(a) Partial density of states (PDOS) of Mn$_3$Ge.  ARPES spectra along the $\Gamma-K$ direction measured with (b) $p-$polarized and (c) $s-$polarized lights overlapped with the DFT band structure (with SOC). Arrows in panel (b) show the ($E, k$) location of various flat bands.    The top panel in (d) shows schematic ARPES measuring geometry with $p-$ and $s-$polarized lights defined with respect to the scattering plane (sp). The bottom panel in (d) shows various Mn $3d$ orbitals projected onto the hexagonal Brillouin zone and oriented w.r.t. the scattering plane. In this measuring geometry, the $p-$polarized light has the even parity and the $s-$polarized light has the odd parity. Similarly, w.r.t. to sp the orbital characters d$_{xz}$, d$_{x^2-y^2}$, and d$_{z^2}$ have the even parity, while d$_{xy}$ and d$_{yz}$ have the odd parity. (e) Orbital-resolved DFT band structure (without SOC) calculated for Mn $3d$ orbitals,  $d_{z^2}$, $d_{xz}$, $d_{x^2-y^2}$, $d_{xy}$, and $d_{yz}$.}
	\label{fig2}
\end{figure*}

Mn$_3$Ge crystallizes in the hexagonal structure as shown in Fig.~\ref{fig1}(a) with a nonsymmorphic space group of $P6_3/mmc$ (No. 194). One unit cell consists of two layers of Mn$_3$Ge, while each layer consists of three Mn and one Ge atoms. The two layers of Mn$_3$Ge stacked along the $\it{c}$-axis are connected by inversion symmetry. The three Mn atoms in a layer form a kagome lattice consisting of triangles and hexagons with the Ge atom positioned at the center of the hexagon~\cite{Kübler_2017, chen2021anomalous}. Fig.~\ref{fig1}(b) shows the hexagonal Brillouin zone (BZ) along with various high symmetry points located at their respective positions.
Fig.~\ref{fig1}(c) shows an in-plane Fermi surface (FS) map measured using $p-$polarized light with a photon energy of $h\nu$=120 eV. Later, we will show that the 120 eV photon energy extracts the electronic bands from $k_z$$\approx$4.0 $\pi/c$ plane, i.e., the $\Gamma KM$ plane of the Brillouin zone.

From Fig. \ref{fig1}(c), we can notice that the FS map has a hexagonal symmetry that is consistent with the hexagonal crystal structure. In the figure, the Fermi pockets on the FS map are shown by the white-colored elliptical shapes. It is to be noted that since multiple bands cross the Fermi level at $K$ and $M$ points, we observe spectral intensities at both $K$ and $M$ points surrounding the $\Gamma$ point as shown in Fig. \ref{fig1}(c). The large number of bands at both $K$ and $M$ points makes it challenging to resolve distinct energy pockets at the Fermi level. Therefore, what may appear as energy pockets in Fig. \ref{fig1}(c) are largely influenced by the matrix element effects, leading to asymmetric spectral intensities. Fig. \ref{fig1}(d) shows a constant energy map taken at 100 meV below the Fermi level (E$_F$). The constant energy map taken at E$_F$-100 meV is consistent with the Fermi surface of Mn$_3$Sn \cite{kuroda2017,PhysRevB.106.205103}, where we  observe six high-intensity Fermi pockets at six $M$ points of the BZ. Moreover, the size of the Fermi pockets decreases as we go from $E_F$ to E$_F$-100 meV, suggesting that the Fermi pockets enclosed in the elliptical shape are of the electron-type. This interpretation is consistent with the EDM shown in Fig. \ref{fig1}(g) from which we can notice four electron-like band dispersions crossing the Fermi level.


For a better understanding on the nature of Fermi pockets, energy distribution maps (EDMs) and their second derivative with respect to momentum along the $\Gamma-K$, $\Gamma-M$, and $M-K$ high symmetry directions are shown in Figs.~\ref{fig1}(e),~\ref{fig1}(f), and~\ref{fig1}(g), respectively. From these EDMs, we notice several well-dispersive electronic bands crossing $E_F$ at high symmetry points $M$ and $K$, while we could not find any band crossing the $E_F$ at the high symmetry point $\Gamma$.

\begin{figure*}[ht]
	\includegraphics[width=0.8\linewidth]{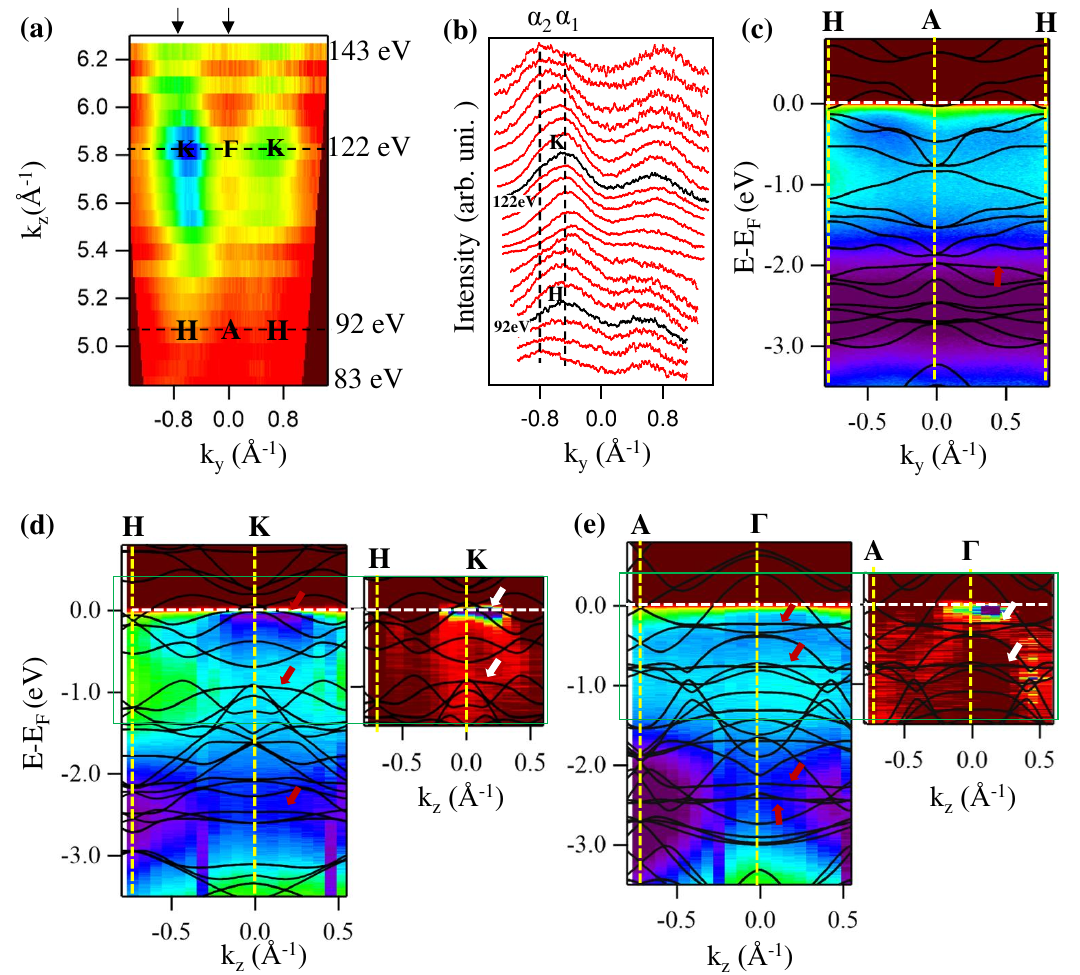}
	\caption{(a) Out-of-plane ($k_z$) Fermi surface map of Mn$_3$Ge. (b) Stacked momentum dispersive curves (MDCs)  taken at the Fermi level from the EDMs measured using the photon energies between 83 and 143 eV.   (c), (d), and (e) EDMs taken along the $A-H$, $K-H$, and $\Gamma-A$ high symmetry directions, respectively. The EDMs shown in (d) and (e) are taken along the cuts shown by the down-arrows in (a). The overlapped black-solid curves in (c)–(e) are the DFT band structure. The arrows in (c)-(e) show the ($E, k$) location of several flat bands. The right panel of (d)-(e) represents the second derivative of the EDMs shown in the left. The green boxes mark the energy range of the EDMs, over which the second derivative was performed.}
	\label{fig3}
\end{figure*}

The electronic band structure calculated using the DFT is overlapped on the EDMs as shown in Figs.~\ref{fig1}(e)-(g). It is to be noted that the Fermi level of the DFT band structure is shifted by 180 meV towards higher binding energy to match the experimental band dispersions, as the Mn deficiency gives a net effect of hole doping into the system. Nevertheless, we find a qualitative agreement between the DFT band structure and the ARPES data for all high symmetry directions, as can be seen from Figs.~\ref{fig1}(e)-(g). An earlier report on Mn$_3$Sn showed the presence of strong electronic correlations, as they had to perform a band renormalization by a factor of five to the DFT band structure in order to obtain a good overlap with ARPES data~\cite{kuroda2017}. In our case, we renormalized the DFT band structure by a factor of 1.18 to match our ARPES data, suggesting that the electronic correlations in Mn$_3$Ge are relatively weaker compared to Mn$_3$Sn. This observation is consistent with an earlier theoretical study, which says that Mn$_3$Ge has a wider bandwidth leading to a smaller band renormalization of a factor of 2~\cite{chen2021anomalous}.

Fig.~\ref{fig2}(a) shows the partial density of states (PDOS) calculated using the DFT. From Fig.~\ref{fig2}(a) it is evident that the Mn $3d$ orbital characters dominate the valence band structure, particularly near $E_F$ only the Mn $3d$ orbitals are present. ARPES spectra collected along the $\Gamma-K$ direction with $p-$ and $s-$polarized lights are displayed in Figs.~\ref{fig2}(b) and~\ref{fig2}(c), respectively. As per the measuring geometry shown Fig.~\ref{fig2}(d), the $p-$polarized light extracts the even parity Mn $3d$ orbital characters like d$_{xz}$, d$_{x^2-y^2}$, and d$_{z^2}$ and the $s-$polarized light extracts the odd parity orbital characters like d$_{xy}$ and d$_{yz}$ in the vicinity of Fermi level. Orbital-resolved band structures for all Mn $3d$ characters are shown in Fig.~\ref{fig2}(e). By closely comparing the ARPES data shown in Figs.~\ref{fig2}(b) and~\ref{fig2}(c) with DFT band structure shown in Fig.~\ref{fig2}(e), we can qualitatively identify that the $p-$polarized light predominantly extracts the d$_{z^2}$ and d$_{x^2-y^2}$ orbitals, while the $s-$polarized light predominantly extracts the d$_{yz}$ orbital character. Moreover, Mn$_3$Ge is known to possess flat bands due to the localization of the Mn $3d$ electrons within the kagome lattice~\cite{PhysRevLett.114.245503, kuroda2017}. In agreement with that, we also observe flat bands around the $K$-point (shown by the arrows) from the DFT calculations. A hint of the flat band feature is also visible near the Fermi level in the EDM taken along the $\Gamma-K$ direction measured with $p-$polarized [see Fig.~\ref{fig2}(b)].

Fig.~\ref{fig3}(a) shows out-of-plane ($k_z$) Fermi surface map of Mn$_3$Ge measured by varying the photon energy between 83 and 143 eV with an energy interval of 3 eV using $p-$polarized light. Using an inner potential of $V_0$=12.5$\pm$0.5 eV, we identified that the photon energy 122 eV extracts the electronic states from $k_z$=4.0 $\pi/c$ (5.84 $\AA^{-1}$) plane and the photon energy 92 eV extracts the states from $k_z$=3.5 $\pi/c$ (5.12 $\AA^{-1}$) plane, following the relation $k_{z} = \sqrt{\frac{2m_e}{\hbar ^2} [E_{kin} cos^2\theta+V_0]}$. Here, $m_e$ is the electron mass and the lattice constant $\it{c}$=4.305 $\AA$.  Therefore, the ARPES data shown in Fig.~\ref{fig1} measured with 120 eV of photon energy are nearly extracted from $k_z$=4.0 $\pi/c$ ($\Gamma KM$) plane. Fig.~\ref{fig3}(b) depicts stacked momentum dispersive curves (MDC) taken at $E_F$ from the EDMs measured using the photon energies between 83 and 143 eV. From the MDCs,  we can identity two bands $\alpha_1$ and $\alpha_2$ at around $K$ point which are hardly dispersing in going from $K$ to $H$ except that the $\alpha_1$ and $\alpha_2$ band intensities vary with photon energy due to the matrix element effects.

Next, the EDM shown in Fig.~\ref{fig3}(c) is measured with a photon energy of 92 eV corresponding to the band dispersion along the $A-H$ high symmetry direction. As can be seen from Fig.~\ref{fig3}(c), the overlapped DFT band structure of the $A-H$ direction quantitatively matches the experimental band dispersions, supporting the interpretation that the 92 eV photon energy extracts the electronic states from the $AHL$ plane. Figs.~\ref{fig3}(d) and~\ref{fig3}(e) show the EDMs taken along the $H-K$ and $A-\Gamma$ directions, respectively. Again, the calculated band structure is qualitatively overlapped on these $k_z$ dependent EDMs. Particularly, we noticed a flat band at around -0.23 eV below the Fermi level from the EDM of the $A-\Gamma$ direction [see Fig.~\ref{fig3}(e)] and several other flat bands are also noticed far below the Fermi level (-2.5 to -3 eV) from the EDMs of the $H-K$ and $A-H$ directions. These experimental observations are consistent with our DFT calculations. It is important to note that despite Mn$_3$Ge displays relatively weaker electronic correlations it also features flat bands. This intriguing observation can be attributed to the kagome lattice network in Mn$_3$Ge rather than to the strong electronic correlations.  In kagome systems the electrons confine locally through the destructive quantum interference, resulting the formation of flat bands. This has been observed in the other kagome systems such as in CoSn~\cite{liu2020orbital} and GdV$_6$Sn$_6$~\cite{ding2023kagome} which also hosts weaker electronic correlations.

Overall, we find a qualitative agreement between the experimental and DFT band structures of Mn$_3$Ge. However, we were unable to detect the Weyl points and the surface Fermi arcs from our ARPES measurements as they are predicted above the Fermi level~\cite{yang2017topological,chen2021anomalous}. Our sample has a 2\% of Mn (Mn$_{2.94}$Ge) deficiency compared to the stoichiometric Mn$_3$Ge, leading to a further shift of the experimental Fermi level towards the higher binding energy, making it difficult to capture the predicted dispersive surface Fermi arcs. To experimentally visualize the Weyl points in Mn$_3$Ge using ARPES, we need Mn$_3$Ge single crystals without disorder. It is worth noting that, so far, the Mn$_3$Ge single crystals were synthesized either with excess or deficient Mn. As a result, the ARPES data suffers from bad energy and momentum resolutions due to the disorder. Additionally, it is essential to note that the Weyl points appear above the Fermi level, which falls outside the detection range of ARPES. To overcome this limitation, the system needs to be doped with electrons in order to shift the Fermi level towards higher kinetic energy to reach the Weyl points. Further,  chiral anomaly is a crucial phenomenon that serves as compelling evidence for the Weyl semimetallic phase. This phenomenon can be detected through the angle-dependent magnetotransport measurements~\cite{Zhang2016}. Thus, the investigation of chiral anomaly in Mn$_3$Ge can offer robust confirmation of the existence of Weyl nodes in this system. In addition, unlike Mn$_3$Sn which shows band renormalization by a factor of five, we do not observe significant band renormalization in the experimental band structure of Mn$_3$Ge when compared with the DFT calculations.  Further, the experimental band structure of this study is also broadened similar to Mn$_3$Sn despite weaker electronic correlations~\cite{chen2021anomalous}. We think, the defects in the case of Mn deficient Mn$_3$Ge (our study) or the Mn impurities in the case of Mn excess Mn$_3$Sn are responsible for the broadened ARPES data rather than the electronic correlations.

\section{Conclusions}
In conclusion, we have systematically investigated the in-plane and out-of-plane electronic band structure of the non-collinear antiferromagnetic Weyl semimetal Mn$_3$Ge using ARPES and DFT calculations. Overall, the ARPES data match qualitatively with the DFT band structure near the Fermi level. Consistent with DFT predictions, we could observe flat bands near and well below the Fermi level. Calculated PDOS suggests that the Mn $3d$ orbital characters dominate the valence band structure, particularly near the Fermi level. With the help of orbital-resolved band structure calculations, we identify various Mn $3d$ orbital contributions to the experimentally obtained band structure. We suggest relatively weaker electronic correlations in Mn$_3$Ge compared to Mn$_3$Sn.

\section{Acknowledgement}

S.C. and S. G. acknowledge the University Grants Commission (UGC), India, for the Ph.D. fellowship. This research has made use of the Technical Research Centre (TRC) instrument facilities of the S. N. Bose National Centre for Basic Sciences, established under the TRC project of the Department of Science and Technology (DST), Govt. of India. A.B. is supported by the Prime Minister's Research Fellowship (PMRF). A. N. thanks the Indian Institute of Science for a start-up grant. S.T. thanks the financial support provided by UGC-DAE CSR through grant no. CRS/2021-22/01/373. S.T. thanks the Science and Engineering Research Board (SERB), Department of Science and Technology (DST), India, for the financial support through grant no. SRG/2020/000393.

\section*{References}
\bibliographystyle{iopart-num}
\bibliography{Mn3Ge}

\end{document}